\documentclass[pdflatex,sn-mathphys-num]{sn-jnl}

\usepackage{graphicx}%
\usepackage{multirow}%
\usepackage{amsmath,amssymb,amsfonts}%
\usepackage{amsthm}%
\usepackage{mathrsfs}%
\usepackage[title]{appendix}%
\usepackage{xcolor}%
\usepackage{textcomp}%
\usepackage{manyfoot}%
\usepackage{booktabs}%
\usepackage{algorithm}%
\usepackage{algorithmicx}%
\usepackage{algpseudocode}%
\usepackage{listings}%
\usepackage{url}
\usepackage{hyperref}

\theoremstyle{thmstyleone}%
\theoremstyle{thmstyletwo}%

\theoremstyle{thmstylethree}%

\begin{document}

\title[sshafe]{SSHafe: A Real-Time SSH Brute Force Attack Detection and Novel Credential Rotation Standard}


\author*[1,3]{\fnm{Aditya} \sur{Mitra}}\email{aditya.mitra@stu.khas.edu.tr}

\author[1,3]{\fnm{Amar Kumar} \sur{Mandal}}\email{amarkumar.mandal@stu.khas.edu.tr}

\author[1,3]{\fnm{Amaan Rais} \sur{Shah}}\email{amaanshah@stu.khas.edu.tr}

\author[1,3]{\fnm{Iqra} \sur{Naz}}\email{iqra.naz@stu.khas.edu.tr}

\author[2,3]{\fnm{E. Fatih} \sur{Yetkin}}\email{fatih.yetkin@khas.edu.tr}

\author[2,3]{\fnm{Tuğçe} \sur{Ballı}}\email{tugce.balli@khas.edu.tr}

\affil*[1]{\orgdiv{CyberMACS}}

\affil[2]{\orgdiv{Department of Management Information Systems}}

\affil[3]{\orgdiv{Kadir Has University}, \orgaddress{\street{Cibali}, \city{Istanbul}, \postcode{34083}, \country{Turkey}}}


\abstract{SSH remains a critical yet heavily targeted protocol for remote system administration, with password‑based authentication exposing servers to large‑scale brute‑force, dictionary, and credential‑spray attacks. Existing rule‑based defences such as Fail2Ban fail to detect slow, distributed, or threshold‑aware adversaries, while conventional account‑recovery mechanisms: email links, OTPs, and out‑of‑band verification introduce additional vulnerabilities including phishing, session hijacking, and weak authentication binding. This work presents SSHafe, a real‑time SSH brute‑force detection and mitigation system that combines time‑series feature engineering with a lightweight LightGBM classifier to identify attack patterns directly from system authentication logs. A multi‑scale sliding‑window approach extracts behavioural features such as attempt rates, inter‑arrival times, failure ratios, and username diversity, enabling the model to achieve a detection accuracy of 99.96\% on benchmark data and strong performance on unlabeled real‑world traffic. Upon detecting an attack, SSHafe automatically blocks the targeted user account and delivers an SSH banner guiding legitimate users to a novel passkey‑based password‑rotation workflow. The proposed novel password reset standard performs authentication and password update in a single cryptographically bound flow, eliminating the need for sessions, cookies, OTPs, or email‑based verification, and mitigating phishing, session hijacking, CSRF, and replay attacks. Experiments on an Azure VM and live adversarial traffic demonstrate that SSHafe can identify and suppress brute‑force activity within ten seconds, preventing account compromise even with weak credentials. By unifying real‑time detection with a secure, sessionless credential‑rotation mechanism, SSHafe provides a robust, end‑to‑end framework for defending SSH services against modern automated attack vectors.}




\maketitle

\section{Introduction}\label{sec1}

The Secure Shell (SSH) protocol is the backbone of remote server administration, used daily across cloud platforms, enterprise infrastructure, and embedded systems worldwide. SSH offers two authentication methods: public key and password. Public keys are cryptographically stronger, but they come with a practical catch: one can only log in from a machine that already holds one's private key. Passwords, on the other hand, let one connect from anywhere, which is why they remain the more common choice despite the security trade-off. That trade-off is significant: passwords are a natural target for brute-force and dictionary attacks, in which automated tools cycle through thousands of credentials per second until one works, granting the attacker immediate access to the system. The standard defence, often implemented by tools like Fail2Ban, is to block an IP after too many failed login attempts, which sounds reasonable. However, it is easily defeated by attackers who simply slow down or spread their attempts across multiple addresses, staying just below the detection threshold \cite{10.1007/978-3-030-58201-2_4}. This gap between real-world attack behavior and what rule-based systems can detect motivates the present work.

Password brute-force attacks are well-characterised zero-knowledge threats \cite{9913204}, yet existing account recovery mechanisms: email magic links, one-time passwords, and out-of-band SMS codes introduce secondary vulnerabilities including session hijacking and phishing \cite{9020194}. A fully effective response, therefore, requires not only reliable detection but also a cryptographically robust, end-to-end credential rotation mechanism.

This study presents SSHafe, a real-time SSH brute-force detection and mitigation system. A machine learning model monitors authentication logs and, upon detecting an attack, immediately blocks the targeted account and alerts the user via an SSH banner with a link to a password reset portal. The user can then securely rotate their credentials through a novel passkey-based reset workflow, which binds authentication and password change into a single phishing-resistant flow, eliminating the vulnerabilities inherent in conventional recovery mechanisms such as OTPs and email magic links.

The main contributions of this study are:
\begin{itemize}
    \item A machine Learning time series model to detect SSH brute force and password spray attacks.
    \item Dynamic configuration of SSH to block accounts facing the brute force attacks.
    \item A novel cryptographic approach for password rotation for the blocked accounts using FIDO2 Passkey standard.
\end{itemize}

\section{Literature Review}\label{sec2}

The intrusion detection problem has remained one of the prominent issues in network security, and the detection method has evolved from simple rule-based to machine learning based methods. This also applies to the Secure Shell (SSH) anomaly detection problem, as in this study. A study \cite{10.1007/978-3-030-58201-2_4} provides evidence that Adaptive Boosting (ADA) Boosted random forest detected 40\% more SSH brute force attacks than traditional rule-based methods like Network Measurement Analysis (NEMEA) detection. The false positive rate decreased from 0.94\% to 0.45\% while maintaining the accuracy of 0.94\% \& Area Under Curve (AUC) of 0.998 because Machine Learning (ML) based solutions were able to incorporate protocol-based features like packet length and inter-packet time, which were often overlooked by simple rule-based methods. Another line of study \cite{GONZLEZ2015} evaluated ensemble-based classifiers such as Bagging, boosting, and Rotation Forest on 34 months of honeynet data comprising of 2.6 million packets, which resulted in a reduction of Type-I error (False positive rate) as low as 0.028\% on packet-based analysis and Type-II error (False negative rate) as low as 0.012\% for session-based analysis. Similarly, a study \cite{Wanjau2021} used Convolutional Neural Network (CNN) based approach on the CIC-IDS 2018 dataset \cite{Sharafaldin2018TowardGA} for brute force attack detection, achieving an accuracy of 94.3\%, a precision of 92.5\%, and a recall of 97.8\%, clearly outperforming the classical ML classifiers like Naive Bayes, Logistic Regression, and Support Vector Machine (SVM).
 
Moving on to the time-based machine learning model. A study \cite{Hossain2020} demonstrated that Long Short Term Memory (LSTM) model achieved an accuracy of 99.88\% on the CICIDS2017 dataset \cite{https://doi.org/10.21227/akxq-9v09} for the SSH Brute Force detection task, outperforming a non-temporal-based machine learning model. \cite{Sharma2024} proposed an LSTM based system called Brute Force Secure Shell (BruSSH) for early detection of distributed brute force using sequential login failure counts, on two different datasets, i.e., SSH Traffic Dataset \cite{10.1007/978-3-030-58201-2_4} and PANDAcap \cite{https://doi.org/10.5281/zenodo.3759652}. Their results outperform those of\cite{10.1007/978-3-030-58201-2_4} in key performance metrics, achieving a recall of 99.7, a precision of 99.98, and an F1-score of 99.89. Another study \cite{Dananjaya2025} compared three different temporal-based ML models, LSTM, Gated Recurrent Unit (GRU), and 1D CNN, on a synthetic dataset, where it was found that GRU achieved the highest F1 score and accuracy amongst all. Similarly, Lee and Lee \cite{Lee2022} combined feature engineering (inter-packet arrival time) with data augmentation (WGAN-GP) to improve the F1-score of SSH detection, achieving an impressive score of 0.999. All these SSH-specific collective studies demonstrate that the ML-based approach is substantially superior to classical rule-based approaches, although direct comparison between time-series specific architecture and well-engineered ensemble methods on identical SSH datasets remains unexplored.

The application of time series analysis is not just limited to SSH but also widely used in intrusion detection. The classical AutoRegressive Integrated Moving Average (ARIMA) model is used in volume anomalies detection \cite{ZareMoayedi2008}, whereas the autoregressive part is with sequential hypothesis testing to identify abrupt changes \cite{QingtaoWu2005} in network behaviour. A study \cite{Freeman2026} proposed a hybrid framework that combines the ARIMA model with a decision tree classifier. They used an ARIMA pipeline to filter out noisy traffic and reduced the packet rows from  30,000 to just 400, making the job of downstream classifier easy to identify attacks, improving the F1-score to 99.71\% while retaining 95.8\% attacks in the filtered rows. In contrast, a matrix profile algorithm (unsupervised learning) achieved an accuracy of 99.6\% \& F1-score of 99.8\% on UNSW-NB15 \cite{https://doi.org/10.26190/5d7ac5b1e8485} and CICIOT2023 dataset \cite{AlMazrawe2025}.

On the other hand, deep temporal models, particularly the LSTM variant, have shown strong performance on the canonical intrusion benchmark, often achieving 99\% accuracy on the NSL-KDD \cite{https://doi.org/10.23721/100/1478792} and CICIDS2017 datasets \cite{https://doi.org/10.21227/akxq-9v09} , \cite{Uwazie2024}. A hybrid approach, such as CNN-LSTM, which combines local pattern extraction with temporal features, has achieved 99.78\% accuracy on the NSL-KDD \cite{Udurume2024} dataset. Similarly, studies by Q. Zhu \cite{Zhu2025} use a graph-based neural network approach, which models relational interaction between network entities and is reported to have achieved 99.99\% accuracy on the benchmark dataset.

One of the most common feature engineering techniques for network data containing logs is the sliding window technique, which extracts important feature vectors necessary for a supervised machine learning model by sliding over the datasets \cite{ZareMoayedi2008}, \cite{QingtaoWu2005}, \cite{Freeman2026}. Features such as attempt rates, failure rates, unique user count, and Inter Arrival Time (IAT) means are computed over multi-scale windows, which help capture different attack patterns from intense to prolonged anomalous behavior \cite{10.1007/978-3-030-58201-2_4}, \cite{QingtaoWu2005}. Several studies support the claim that aggregation-based feature engineering techniques improve classifiers' sensitivity to detect an attack behavior \cite{Nguyen2023}, \cite{Ali2025}.

Tree-based ensemble methods, such as random forests, outperform other models and architectures when working with a structured intrusion detection task by achieving a competitive accuracy of 99.88\% and an F1-score of 97.6\% while also offering high interpretability and low computational cost during training \cite{Ali2025}. In contrast, gradient boosting frameworks such as Light Gradient Boosting Machine (LightGBM) are more suitable for heterogeneous tabular features due to their ability to work with mixed feature vectors and learn non-linear decision boundaries \cite{Freeman2026} \cite{Ali2025}.

The existing literature reveals that most work focuses either on deep learning based architecture, such as LSTM \cite{Hossain2020}, \cite{Sharma2024}, \cite{Dananjaya2025}, and CNN \cite{Wanjau2021}, or classical ML classifiers trained on non-temporal per connection-based features \cite{10.1007/978-3-030-58201-2_4}. In addition to it, the majority of prior work did analysis using network-level packet data, such as flow records \cite{10.1007/978-3-030-58201-2_4}, \cite{GONZLEZ2015}, \cite{Wanjau2021}, \cite{Lee2022}, rather than application logs from the system, which provide direct access to ground truth like login outcomes, usernames, and authentication methods. Moreover, the existing literature predominantly focuses on detection in isolation, with little attention given to real-time mitigation strategies. 

The proposed work addressed this gap by combining time-series analysis of the authentication log (/var/log/auth.log) with a lightweight machine learning model as a downstream classifier to detect attacks before they escalate. In addition, our proactive approach focuses on a real-time mitigation strategy that detects attacks and sends an alert to the user to take action and secure their account before the attacker can exploit weak authentication credentials. Table \ref{Tab:modelperformance} compares the performance of the models.

\begin{table}[htbp]
\caption{Performance Metric Comparison}
\label{Tab:modelperformance}
\begin{tabular}{|p{1.5cm}|p{1.2cm}|p{1.2cm}|p{1.2cm}|p{1.2cm}|}
\hline
\textbf{Metric} & \textbf{SSHafe} & \textbf{BruSSH\cite{Sharma2024}} & \textbf{\cite{10.1007/978-3-030-58201-2_4}} & \textbf{CNN \cite{Wanjau2021}} \\
\hline
Recall & \textbf{99.96\%} & 99.70\% & 99.49\% & 97.80\% \\
\hline
Precision & \textbf{100.00\%} & 99.98\% & 99.38\% & 92.50\% \\
\hline
F1-Score & \textbf{99.98\%} & 99.89\% & 99.43\% & 91.80\% \\
\hline
\multicolumn{5}{l}{Note: Bold values indicate the best performance.}
\end{tabular}

\end{table}

One of the most important aspects of a successful attack detection pipeline is the ability to gain insights and take real-time decisions to secure systems, often critical, in the face of attacks from various types of threat actors. For systems involving passwords, the major threats involving zero-knowledge attacks are brute force, dictionary attacks, etc. \cite{Tiwari2023}. And one of the most proactive ways to mitigate such risks is password expiration and rotation. One of the most widely used methods for account recovery after an expired password is using recovery codes or links over emails, SMS-es, other out-of-band factors, etc. A study \cite{Li2022} discussed the vulnerabilities of such account recovery flows and offers a multi-factor Secure Email Account Recovery (SEAR) protocol involving SMS and e-mails. Another study \cite{Sun2012} proposes one-time rotating passwords that are generated on a smartphone app. A study comparing passwordless authentication methods with FIDO2 against other authentication methods \cite{Yusop2025} highlighted the usability of FIDO2 against other account recovery channels. 

However, all these account recovery methods often authenticate the user first over methods like email magic links, passwordless FIDO2, or other factors, and then initiate the password reset flow. This creates a binding issue between the one who authenticated and the one who is entering the new password. This is often dealt with using approaches like session cookies, JWTs, and more. However, a study \cite{281384} shows these cookies and tokens are often not expired or scoped well, which allows further account takeover after recovery. Another study \cite{Innocenti2021} shows Cross-Site Request Forgery (CSRF) and other forms of attacks that may allow attackers to take over the authenticated session during account recovery. To mitigate these issues, this study proposes a novel FIDO2-based password reset flow that performs the authentication and password recovery in a single flow, without involving session cookies, tokens, or other approaches requiring a long-lived authenticated session. Table \ref{Tab:securityref} compares SSHafe Password Rotation flow with other commonly used methods.

\begin{table}[htbp]
\caption{Security Comparison of Authentication Approaches}
\label{Tab:securityref}

\begin{tabular}{|p{1.5cm}|p{1.2cm}|p{1.5cm}|p{1.2cm}|p{1.8cm}|}
\hline
\textbf{Approach} & \textbf{Vulnerable to Email theft} & \textbf{Vulnerable to Brute force or dictionary attacks} & \textbf{Vulnerable to session hijack} & \textbf{Vulnerable to Social engineering} \\
\hline
Email magic link & Yes & No & Yes & Yes (Like CSRF after session is authenticated) \\
\hline
Old password & No & Yes & Yes & Yes (Like Phishing) \\
\hline
OTPs & No & Yes & Yes & Yes (Like Phishing) \\
\hline
\textbf{SSHafe Password Rotation} & \textbf{No} & \textbf{No} & \textbf{No} & \textbf{No} \\
\hline
\end{tabular}
\end{table}

\section{Methodology}\label{sec3}

The proposed study uses a machine learning model trained separately and then deployed on a cloud virtual machine (VM) for mitigating attacks.

\subsection{Training workflow}
A labelled dataset \cite{dananjaya_ssh_2024} on SSH information was used to train a Light Gradient Boosting Machine (LightGBM) model. The dataset had columns timestamp, source IP address, Username, event type, status, and label. The dataset had 90\% Brute force records and 9\% normal or non-anomalous records. While this looks like an unbalanced dataset at first glance, it resembles real-life data. The high count of brute force data can be attributed to the nature of brute force attacks, which, by definition, implies a malicious actor trying random passwords very fast. Hence, the number of failures and brute force labelled data is naturally very high.

However, transforming this data to time domain data per user with a sliding window of 60 seconds effectively normalizes the dataset. This is because a high number of failed attempts to login to a particular user account in a short window of around 60 seconds is represented as a single attempt to brute force.

\begin{figure}[!htbp]
\centerline{\includegraphics[width=\columnwidth]{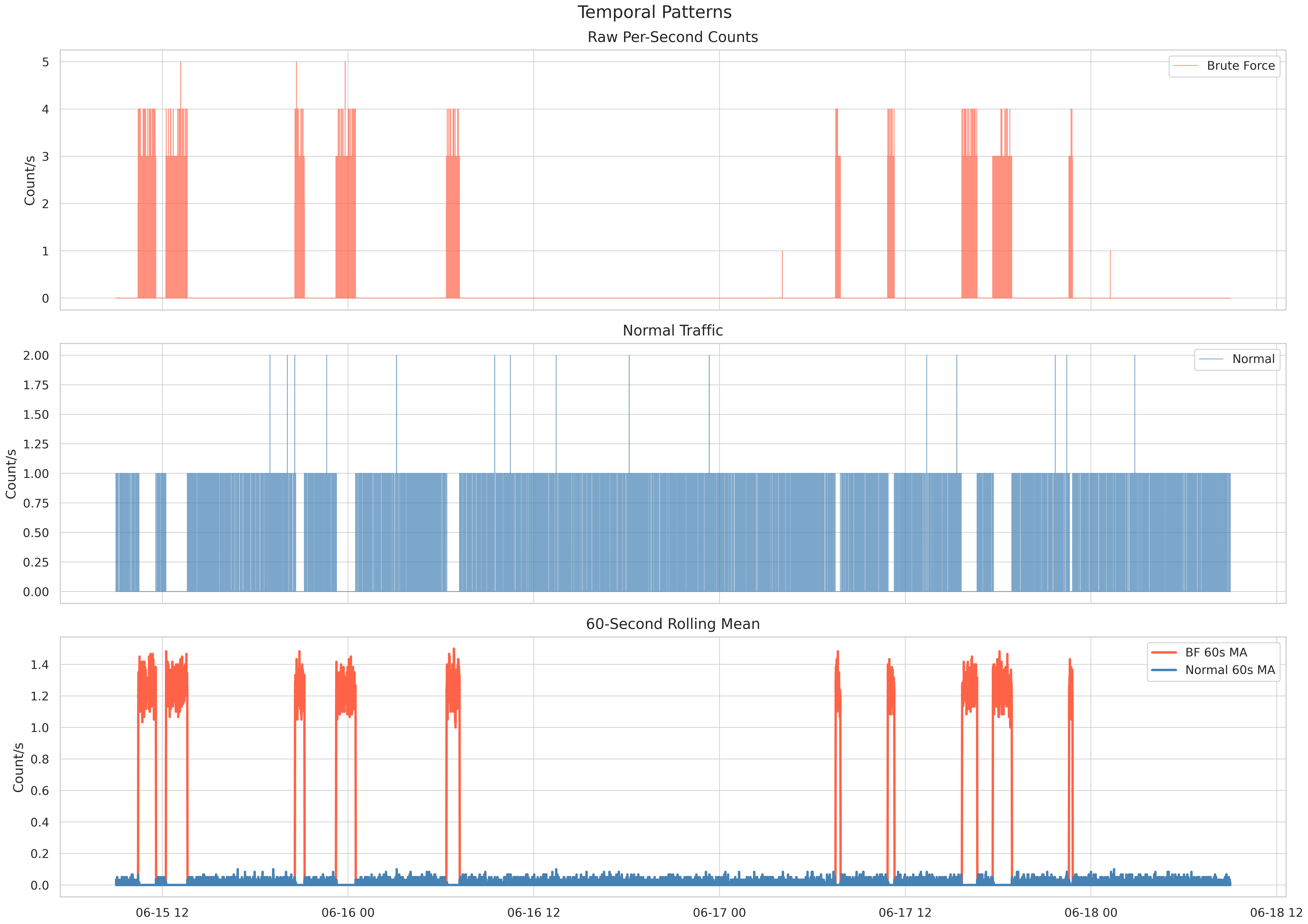}}
\caption{Temporal patterns of SSH attempts}
\label{FIG:temporal}
\end{figure}

An exploratory analysis of the dataset was performed to determine whether it was fit for the purpose of this study. Raw count of both brute force records and normal access records was plotted along with their 60s rolling mean. Brute force attacks showed a high spike in very low time span to around 3 to 5 hits per second. This aligns with the definition of brute force and dictionary attacks where the attacker attempts to spray a high number of random passwords or from a wordlist in a very short amount of time. On the other hand, normal connections usually had only one successful connection or failure per second, with up to a maximum of two per second. This aligns with the general connection pattern where the user takes time to enter the password or does not attempt multiple connections very fast. This creates a very visible baseline for further analysis of the dataset. Figure \ref{FIG:temporal} shows the temporal features of the dataset and the baseline difference between the count of attempts of a particular type (benign and attack) per second. It also shows a 60-second rolling mean for comparison.

\begin{figure}[!htbp]
\centerline{\includegraphics[width=\columnwidth]{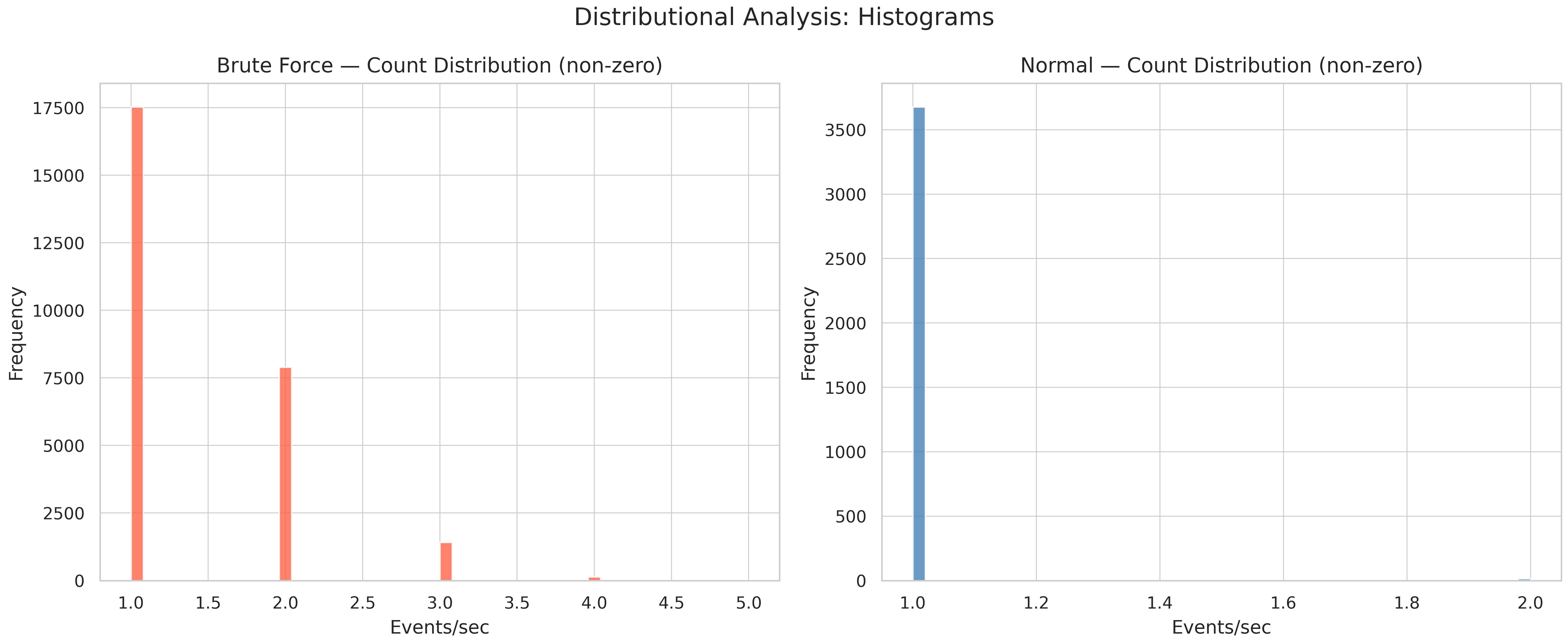}}
\caption{Distribution patterns of SSH attempts}
\label{FIG:distribution}
\end{figure}

A count distribution plot made it clearer. This shows brute force attacks have a wider range of intensity where the brute force attack often reaches 3 to 5 attempts per second, as compared to normal connectivity that is usually only one event a second. Brute force attacks have a high frequency compared to normal connections and are a signature of automated tools or scripts trying to authenticate quickly. Figure \ref{FIG:distribution} shows the distributional patterns of the SSH attempts in the dataset.

Stationarity tests were performed on the dataset. Two tests, Augmented Dickey-Fuller (ADF) and Kwiatkowski-Phillips-Schmidt-Shin (KPSS) tests were performed. The results were contradictory, ADF test showed the dataset was probably stationary, while KPSS showed the dataset was probably non-stationary. This is interpreted as a valid result since the data is stationary. However, the given dataset showed attackers initiating the brute force attack periodically which could have influenced the KPSS model to find out a non-stationary relation. Frequency-domain analysis on the dataset gave no actionable insight due to the absence of any harmonics.

\subsection{Feature Engineering}

Sliding windows of 30, 60 and 300 seconds were used to extract features like attempt rate, failure ratio, unique users, inter arrival time (IAT) mean and covariance, and number of attempts. The 30 second window was plotted for analysis, with the other windows showing similar results.

\begin{figure}[!htbp]
\centerline{\includegraphics[width=\columnwidth]{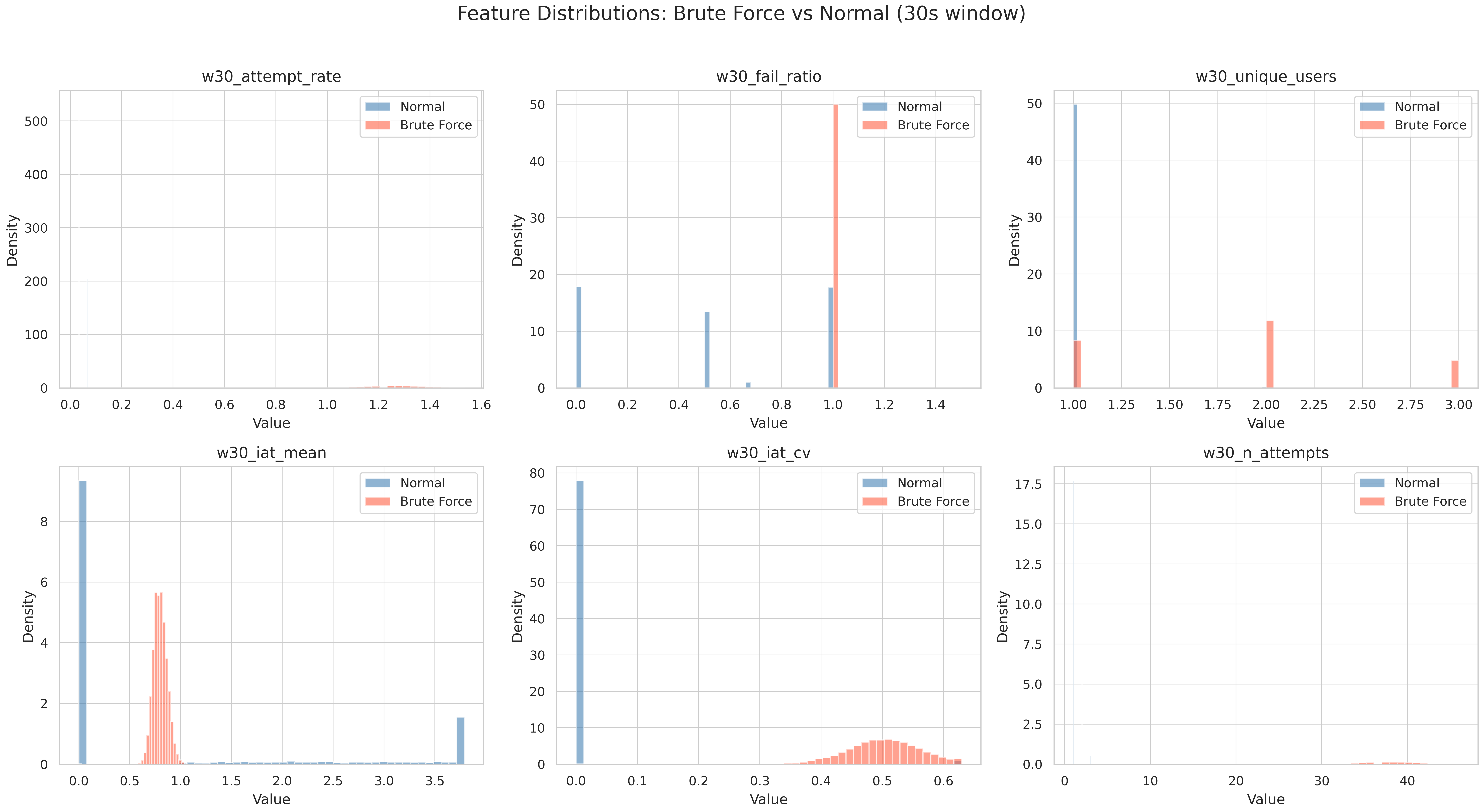}}
\caption{Features extracted from sliding window}
\label{FIG:feature}
\end{figure}

The attempt rate was considerably higher for brute force attempts. The failure ratio density for brute force attacks was around 50 while for normal attempts it was lower. Brute force attempts used higher number of usernames for username spray type attacks. The mean IAT for brute force attacks was considerably lower, and the covariance was higher. The number of brute force attempts in 30 seconds sliding window was around 38, while for normal connections, it was very low, around 1 or 2. Figure \ref{FIG:feature} represents the extracted features over a thirty second sliding window.

The features were considered, and a LightGBM model was trained. The accuracy was 99.96\%, and the model was empirically validated against unlabeled real-world data with domain expert validation, reaching satisfactory performance. The system was found to focus on differentiating burst patterns of failed attempts to identify brute-force attempts.

\begin{figure}[!htbp]
\centerline{\includegraphics[width=\columnwidth]{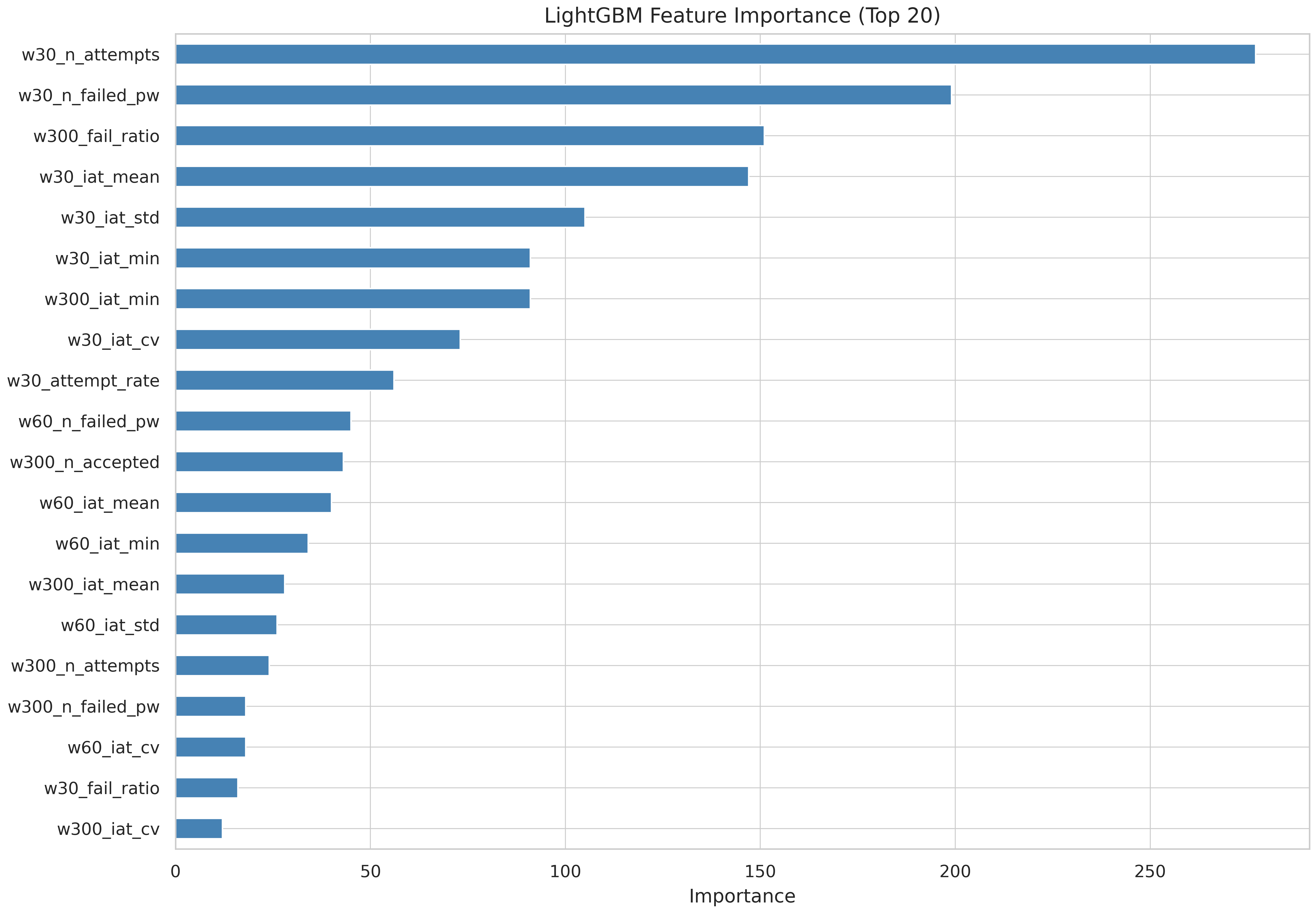}}
\caption{Importance of features as used by LightGBM}
\label{FIG:featureimp}
\end{figure}

The LightGBM feature importance list showed important features that matched empirical validation. The most important features considered by LightGBM were the number of attempts in 30 seconds sliding window, the number of failures in 30 seconds sliding window, the failure ratio in 300 seconds sliding window, and the mean IAT in 30 seconds sliding window. Figure \ref{FIG:featureimp} represents the importance of the features as used by the LightGBM model.

\begin{figure*}[!htbp]
\centerline{\includegraphics[width=\linewidth]{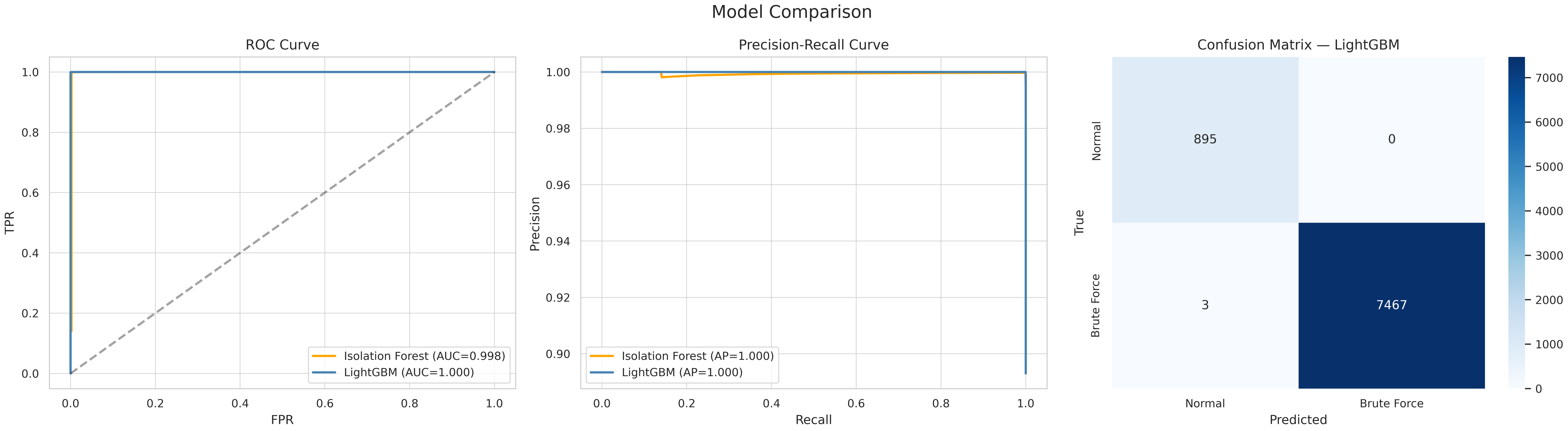}}
\caption{Performance and Confusion matrix}
\label{FIG:confusion}
\end{figure*}

The model reached high accuracy on the test dataset giving minimal false negatives and no false positives. Here false negatives imply the model predicted a request to be benign where it was a brute force request. This is not a problem, since brute force attacks depend on a large number of attempts in a short span of time, even if one is predicted as benign, the next one could be marked as an attack, thus taking the same actions. Figure \ref{FIG:confusion} represents the performance of the model and shows the confusion matrix.

\begin{figure}[!htbp]
\centerline{\includegraphics[width=\columnwidth]{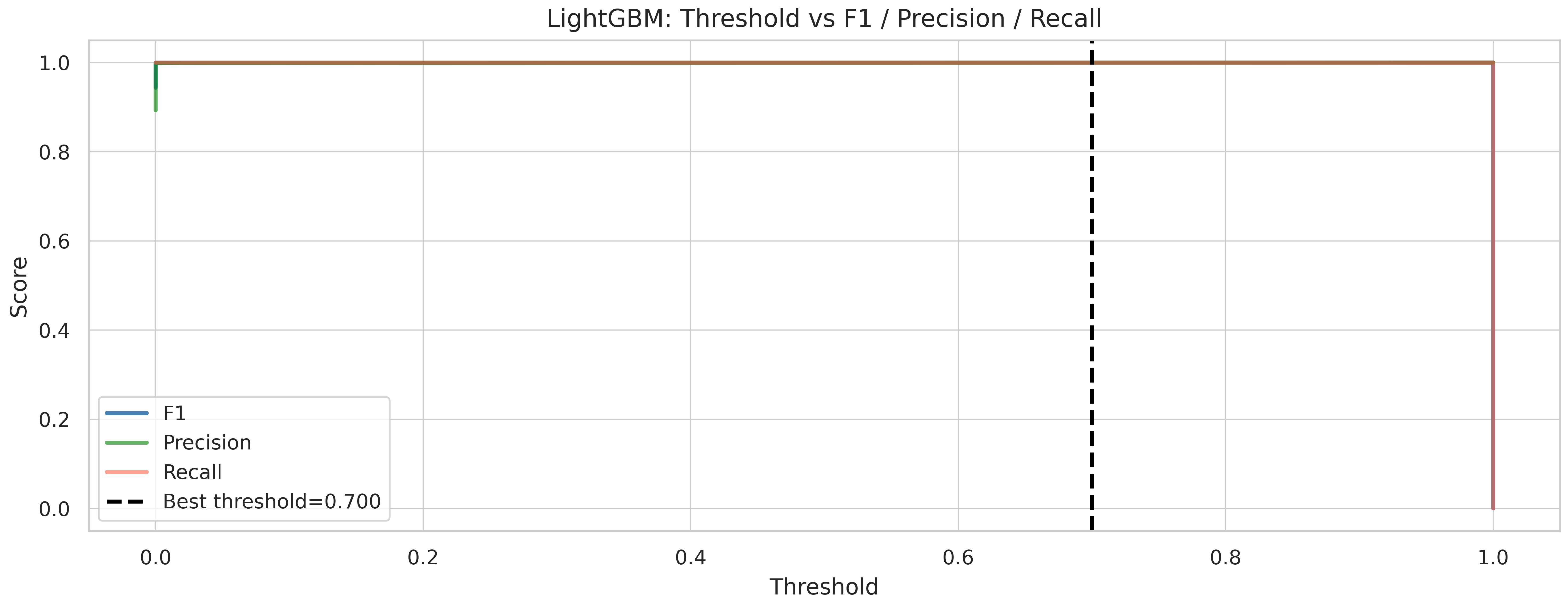}}
\caption{Fine-tuning for best threshold}
\label{FIG:lightgbm}
\end{figure}

The threshold of 0.7 gave the maximum F1 score of 0.9998 and hence was chosen as the best threshold for using the model. The model was fine-tuned to use the best threshold for future prediction. Figure \ref{FIG:lightgbm} shows the threshold of 0.7 giving the highest F1 score.

\begin{figure}[!htbp]
\centerline{\includegraphics[width=\columnwidth]{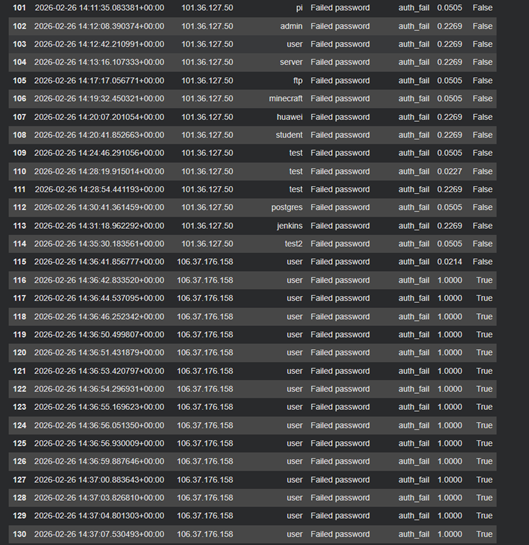}}
\caption{Validation with unlabelled data}
\label{FIG:results}
\end{figure}

A conversion pipeline was developed to read the authentication logs (/var/log/auth.log) from a Linux machine and process it into an unlabeled dataset. Such data from 4 different cloud VMs were used to empirically validate model performance with domain expert knowledge. Figure \ref{FIG:results} shows a section of the validation results of the model with logs from cloud VMs.

The model performance was satisfactory and deemed ready to be deployed for real-world usage.

\subsection{Cross-Validation and Stability Analysis}
To find out the generalization ability and the variance of the proposed detector, we have performed stratified 5-fold cross-validation on the labeled dataset. In each fold, LightGBM was retrained from scratch, and its performance was evaluated on the held-out partition. An unsupervised Isolation Forest trained on the same partition was used as a baseline model against which the performance of LightGBM was measured. Per-fold results are summarized in Table~\ref{Tab:cvfolds}.

\begin{table}[htbp]
\caption{Per-Fold Cross-Validation Results}
\label{Tab:cvfolds}

\begin{tabular}{|c|c|c|c|c|}
\hline
\textbf{Fold} & \textbf{LGB AUC} & \textbf{LGB F1} & \textbf{ISO AUC} & \textbf{ISO F1} \\
\hline
1 & $\mathrm{NaN}$ & 1.0000 & $\mathrm{NaN}$ & 1.0000 \\
\hline
2 & 1.0000 & 0.9998 & 0.9915 & 0.9991 \\
\hline
3 & 1.0000 & 0.9999 & 0.9996 & 0.9968 \\
\hline
4 & 1.0000 & 0.9996 & 0.9993 & 0.9882 \\
\hline
5 & 1.0000 & 0.9999 & 0.9950 & 0.9978 \\
\hline
\multicolumn{5}{l}{\footnotesize $^{*}$AUC $\mathrm{NaN}$: held-out partition contained a single class.}
\end{tabular}
\end{table}

The aggregated statistics in the five folds, including the mean, standard deviation, and the 95\% confidence interval (CI) for each metric, are presented in Table~\ref{Tab:cvstability}. It shows that LightGBM achieved highly stable performance with an F1-score of $0.9998 \pm 0.0001$ and accuracy of $0.9997 \pm 0.0002$. The isolation forest also performed strongly but showed a higher variance ($\text{F1} = 0.9964 \pm 0.0047$). Since precision across all folds remained consistent (mean $= 1.0000$, std $= 0.0000$) for LightGBM, the confidence interval is undefined due to zero variance, indicating that no benign samples were misclassified. The mean recall of $0.9997 \pm 0.0003$ further confirms that very few brute-force events were missed. In contrast, the Isolation Forest baseline produced a wider spread, with the F1-score varying by almost half a percentage point between folds (std $= 0.0047$), reflecting the greater sensitivity of unsupervised anomaly detectors to the composition of each fold.

\begin{table}[htbp]
\caption{Cross-Validation Stability Report (Mean $\pm$ Std, 95\% CI)}
\label{Tab:cvstability}
\setlength{\tabcolsep}{3pt}
\begin{tabular}{|p{1.3cm}|p{1.3cm}|p{2.0cm}|p{1.9cm}|}
\hline
\textbf{Metric} & \textbf{Model} & \textbf{Mean $\pm$ Std} & \textbf{95\% CI} \\
\hline
F1 & LightGBM & $0.9998 \pm 0.0001$ & [0.9997, 1.0000] \\
\hline
Precision & LightGBM & $1.0000 \pm 0.0000$ & [1.0000, 1.0000] \\
\hline
Recall & LightGBM & $0.9997 \pm 0.0003$ & [0.9993, 1.0000] \\
\hline
Accuracy & LightGBM & $0.9997 \pm 0.0002$ & [0.9995, 1.0000] \\
\hline
F1 & IsoForest & $0.9964 \pm 0.0047$ & [0.9905, 1.0000] \\
\hline
Precision & IsoForest & $0.9929 \pm 0.0093$ & [0.9814, 1.0000] \\
\hline
Recall & IsoForest & $1.0000 \pm 0.0001$ & [0.9998, 1.0000] \\
\hline
Accuracy & IsoForest & $0.9942 \pm 0.0068$ & [0.9858, 1.0000] \\
\hline
\end{tabular}
\end{table}

To further quantify the stability of our detector model (i.e., LightGBM), the model was additionally trained on 20 different random seeds ($7, 17, 27, \ldots, 197$) while keeping the feature pipeline fixed. Across all the twenty runs, the model converged to identical metrics of $\text{AUC} = 1.0000$, $\text{F1} = 0.9998$, $\text{Precision} = 1.0000$, and $\text{Recall} = 0.9996$, with no variance between the seeds. This conclude that the decision boundary learned by our LightGBM model is solely dependent on the discriminative temporal features extracted by sliding window rather than stochastic choices made during training which is an important characteristic for the application deployed in security critical infrastructure. Table~\ref{Tab:seedstability} summarises the seed-based stability evaluation.

\begin{table}[htbp]
\caption{Seed-Based Stability (20 Seeds)}
\label{Tab:seedstability}

\begin{tabular}{|l|c|c|}
\hline
\textbf{Metric} & \textbf{Mean} & \textbf{Std.\ Deviation} \\
\hline
AUC & 1.0000 & 0.0000 \\
\hline
F1-Score & 0.9998 & 0.0000 \\
\hline
Precision & 1.0000 & 0.0000 \\
\hline
Recall & 0.9996 & 0.0000 \\
\hline
\end{tabular}
\end{table}

The combined results from cross-fold validation and seed stability experiment provide evidence that our model has achieved high accuracy with minimal variance across all different data split and training run. This consistency makes it well-suited for real-time use on production SSH servers.

\subsection{Attack mitigation}

\begin{figure*}[!htbp]
\centerline{\includegraphics[width=\linewidth]{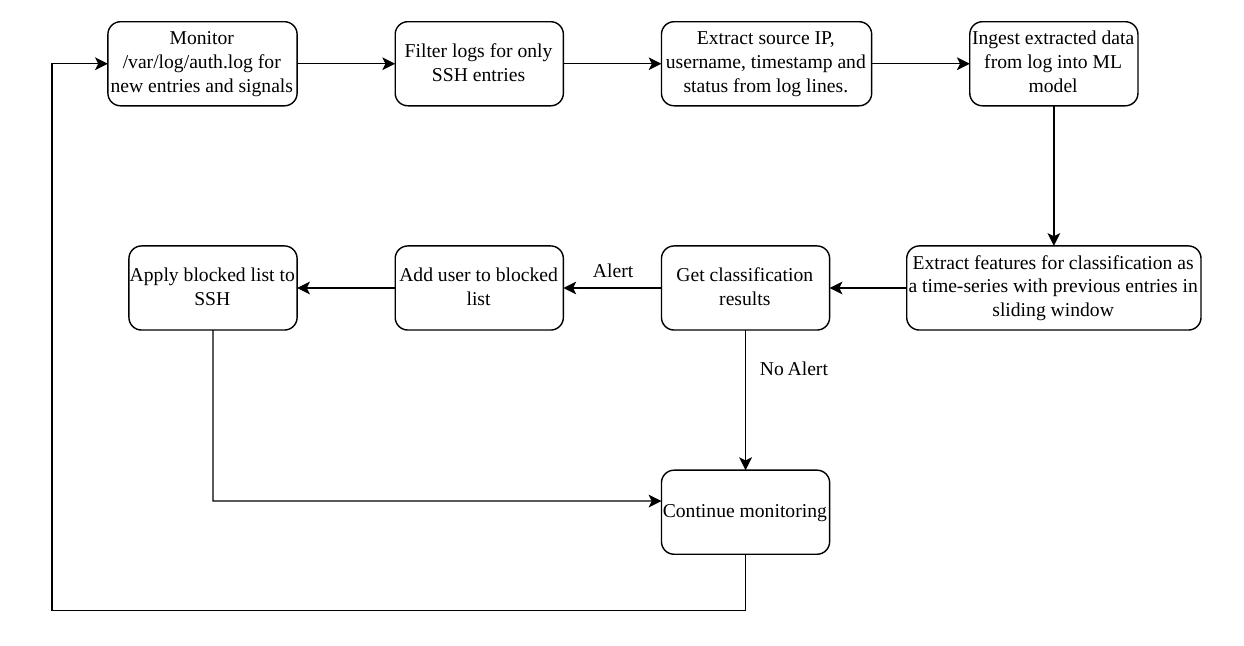}}
\caption{Attack mitigation flow}
\label{FIG:mitigation}
\end{figure*}

The study proposes a Daemon that would monitor the authentication logs and use the machine learning model to detect SSH brute force attacks. The system maintains a list of blocked accounts. Once a brute force attack is detected on a valid user account, the user account is appended to the list. This list uses a ‘Match User’ directive to find the blocked user and disable Password Authentication in SSH configurations. A SSH banner is added for these users showing password authentication was disabled and providing the web link to the password reset portal. For unblocking a user, the username is removed from the list, and the updated configurations are applied to SSH. The user account is automatically unlocked after a successful password reset. Figure \ref{FIG:mitigation} highlights the attack mitigation flow.

\subsection{SSHafe Password Rotation}

\begin{figure*}[!htbp]
\centerline{\includegraphics[width=\linewidth]{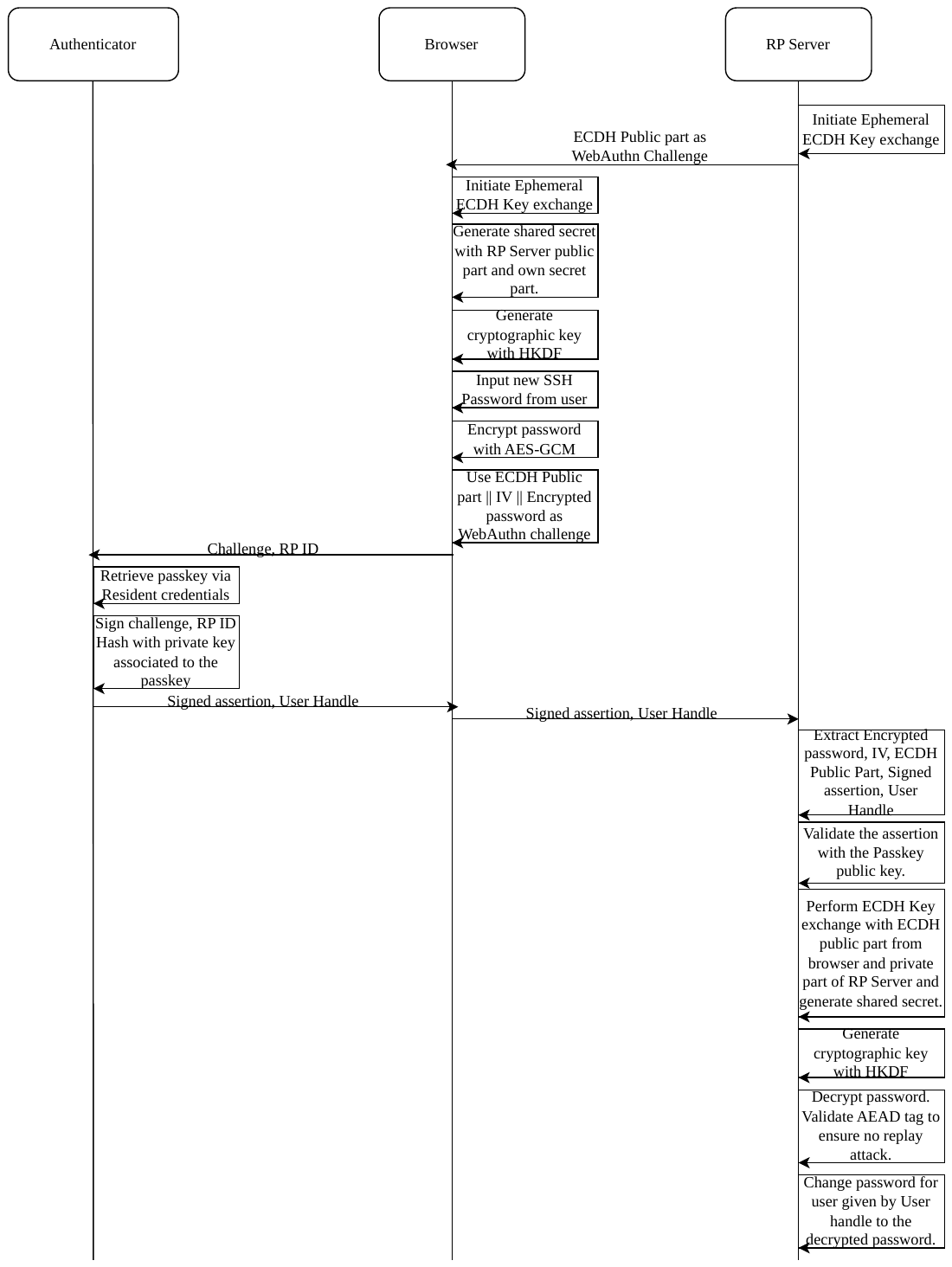}}
\caption{Password Rotation Flow}
\label{FIG:passrot}
\end{figure*}

This study proposes a novel password reset workflow that encapsulates passwords in passkeys to provide non-repudiation and integrity guarantees, while at the same time mitigating vulnerabilities relating to other forms of authentication before password reset, including OTPs, Tokens sent on emails, authentication with out-of-band credentials, and so on. Most password reset flows make authentication and entering password a two-step process, while involving other methods like cookies, JWTs, session storage, etc., to maintain the session. This introduces additional risks like cookie theft, session hijack, and so on.
The proposed password reset workflow relies on phishing-resistant, passwordless authentication factors to carry the password and uses Web Authentication as an authenticated, encrypted channel. When a user account is created or modified, the SSHafe system provides a tool to enroll Passkeys for the account. The tool presents a magic link on the SSH terminal itself to enroll Passkeys with Resident Credentials for the user account. This magic link is tightly scoped to expire in 10 minutes and allows enrollment only for the account it is issued for. This passkey is later used for account recovery. The enrollment is done in-band to prevent attacks on email or other transport mechanism. This magic link is single-use with 10 minute expiry, and the security of it is beyond the scope of this study. Algorithm \ref{Alg:Rotation} represents the password rotation flow and figure \ref{FIG:passrot} shows the steps of SSHafe Password rotation flow.

\begin{algorithm}
\caption{SSHafe: Password Rotation Algorithm}
\label{Alg:Rotation}
\begin{algorithmic}[1]
    \State RP Server initiates an Ephemeral ECDH key pair $(\text{sk}_{rp}, \text{pk}_{rp})$ 
    \State $\text{pk}_{rp}$ is encapsulated as the \texttt{challenge} in a \texttt{PublicKeyCredentialRequestOptions}; \texttt{allowCredentials} is left empty (usernameless flow)
    \State \texttt{PublicKeyCredentialRequestOptions} is transmitted to the client
    \State Client browser generates its own ephemeral ECDH key pair $(\text{sk}_{br}, \text{pk}_{br})$
    \State Browser extracts $\text{pk}_{rp}$ from the challenge and computes shared secret $S \gets \text{ECDH}(\text{sk}_{br}, \text{pk}_{rp})$
    \State Derive encryption key $K \gets \text{HKDF}(S,\ \text{salt} = \text{pk}_{rp})$
    \State Obtain password $P$ from user; sample random IV; compute $C \gets \text{AES-GCM-256}_K(P,\ \text{IV})$
    \State Replace \texttt{challenge} in \texttt{PublicKeyCredentialRequestOptions} with the concatenation $\text{pk}_{br} \| \text{IV} \| C$
    \State Complete the WebAuthn assertion flow using a Discoverable (Resident) Credential with the modified \texttt{PublicKeyCredentialRequestOptions}
    \State Assertion $\mathcal{A}$ is returned to the RP Server
    \State RP Server extracts the modified challenge from \texttt{clientDataJSON} $\in \mathcal{A}$
    \State RP Server extracts \texttt{userHandle} from $\mathcal{A}$ and retrieves the corresponding registered public key $\text{pk}_{user}$
    \State RP Server verifies assertion signature over modified challenge using $\text{pk}_{user}$
    \If{verification fails}
        \State \textbf{abort}
    \EndIf
    \State RP Server parses the modified challenge to recover $\text{pk}_{br}$, IV, and $C$
    \State Compute shared secret $S \gets \text{ECDH}(\text{sk}_{rp}, \text{pk}_{br})$
    \State Derive decryption key $K \gets \text{HKDF}(S,\ \text{salt} = \text{pk}_{rp})$
    \State Decrypt $P \gets \text{AES-GCM-256}_K^{-1}(C,\ \text{IV})$; AEAD tag validation defeats replay attacks
    \If{AEAD tag validation fails}
        \State \textbf{abort}
    \EndIf
    \State Set the password for \texttt{userHandle} to $P$
\end{algorithmic}
\end{algorithm}

\section{Experimental Setup and Validation}
The LightGBM-based machine learning model for this study was trained on a Google Colab notebook, and the model weights were then used on a cloud VM for real-world validation. An Azure B2ats series Virtual Machine with 2 vCPUs and 1 GB of memory with Ubuntu 24.04 LTS operating system was used. This configuration has nearly the least system capabilities in modern days, which further assures that the performance of the model on machines with better configurations will definitely be better than this VM. 

\begin{figure}[!htbp]
\centerline{\includegraphics[width=\columnwidth]{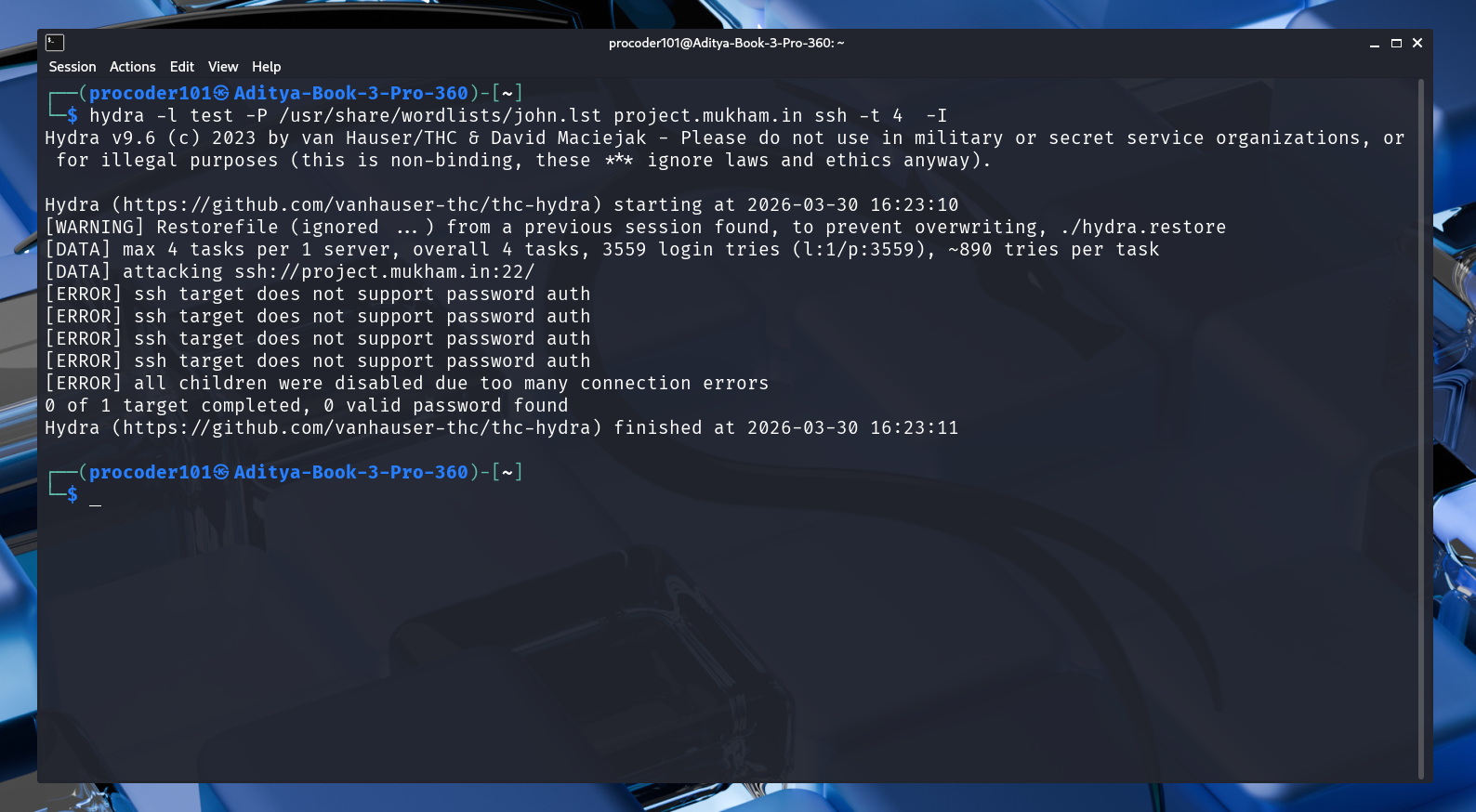}}
\caption{Hydra attack}
\label{FIG:hydra}
\end{figure}

On the other hand, a standard laptop with Kali Linux was used to simulate the attacker. The Hydra tool was used with the Rockyou wordlist, a list of approximately 14.3 million unique passwords. The target VM was able to detect the attacker under 10 seconds in all of 15 tries and disabled the user account. The user account deliberately had a weak password for the tests, yet it could not be taken over by attackers. Figure \ref{FIG:hydra} shows Hydra being used to perform an SSH brute force attack.

This VM was left running on the cloud for two weeks, where it was naturally discovered by real adversarial actors, and they tried to attack the system. The system successfully defended against the attacks from the real adversarial actors, even when both the username and password were relatively weak.

\begin{figure}[!htbp]
\centerline{\includegraphics[width=\columnwidth]{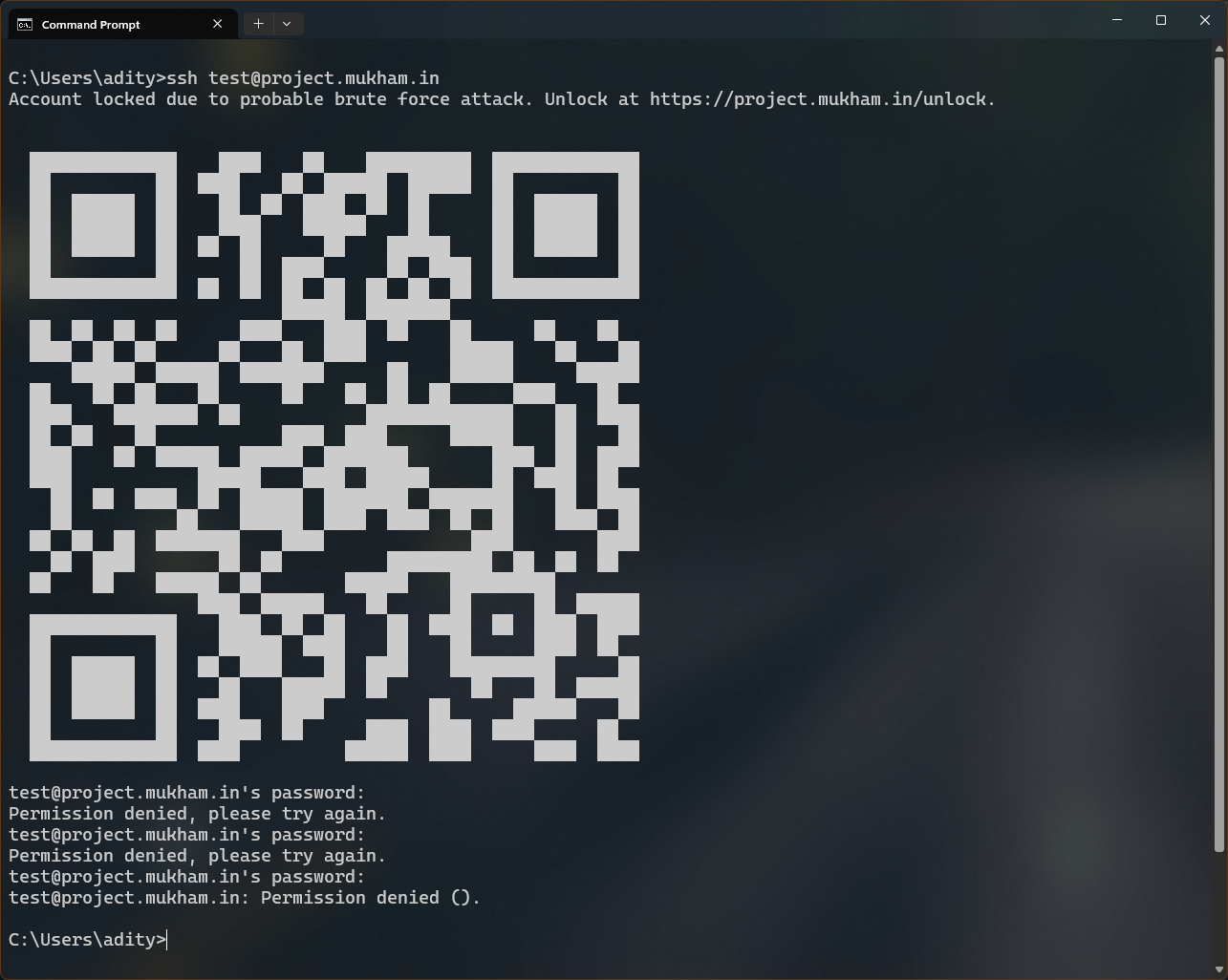}}
\caption{Account blocked message}
\label{FIG:banner}
\end{figure}

The blocked user accounts showed a banner with a link and QR code to the account recovery portal for legitimate users to be able to recover the accounts. Figure \ref{FIG:banner} shows the banner used for blocked accounts.

\begin{figure}[!htbp]
\centerline{\includegraphics[width=\columnwidth]{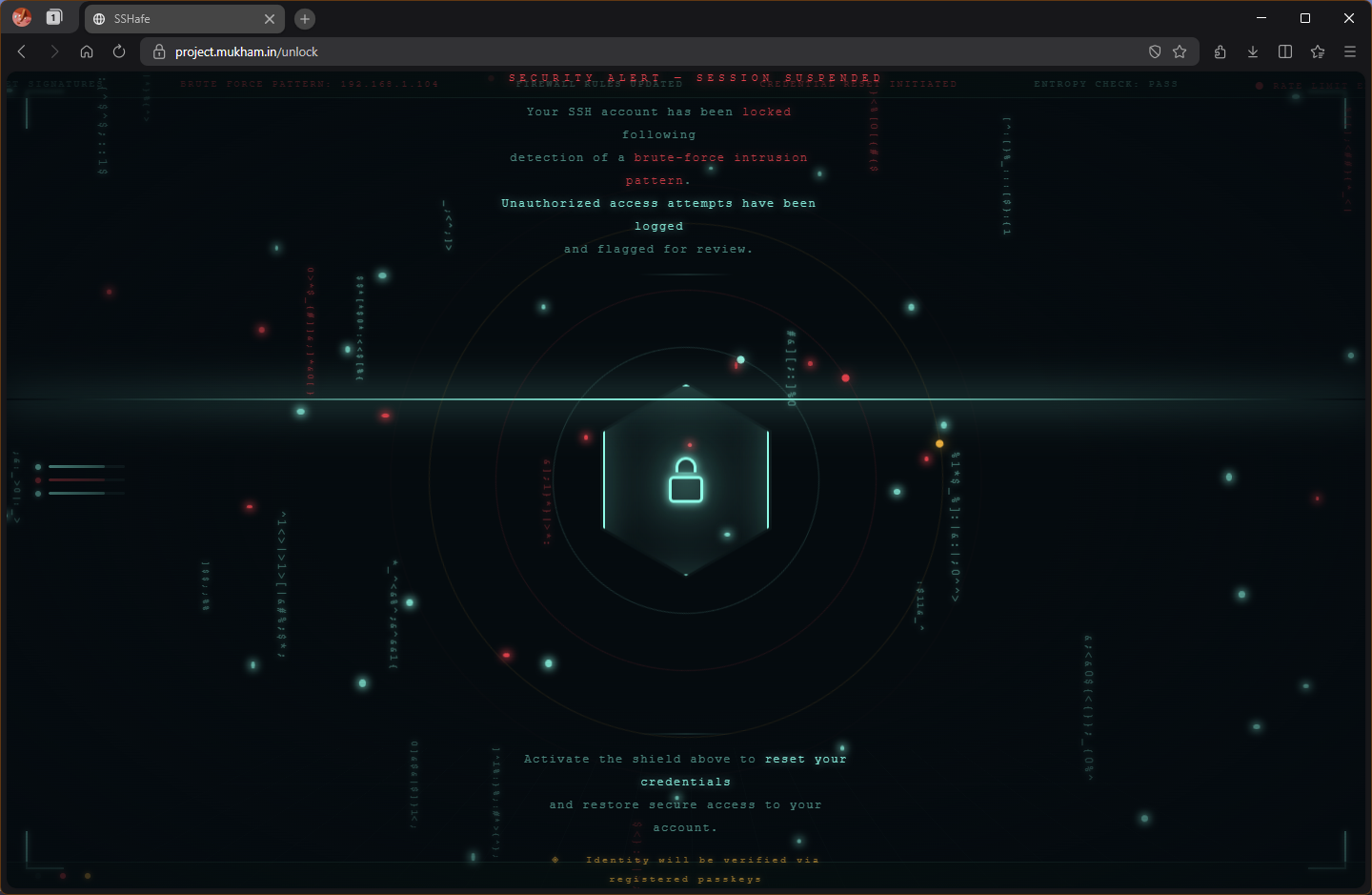}}
\caption{Password rotation homepage}
\label{FIG:homepage}
\end{figure}

\begin{figure}[!htbp]
\centerline{\includegraphics[width=\columnwidth]{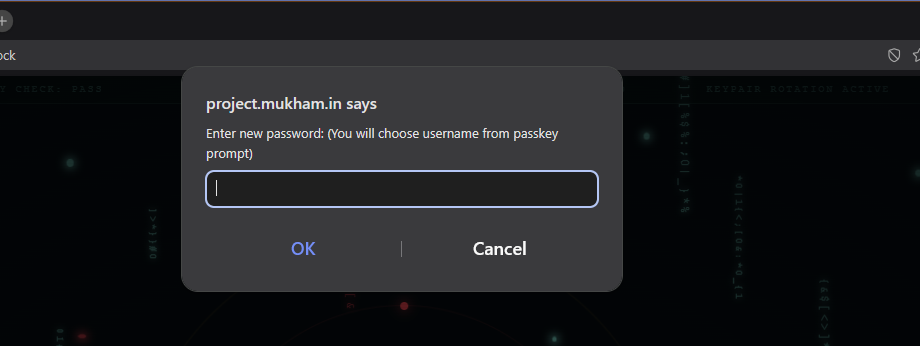}}
\caption{Prompt for new password}
\label{FIG:passprompt}
\end{figure}

\begin{figure}[!htbp]
\centerline{\includegraphics[width=\columnwidth]{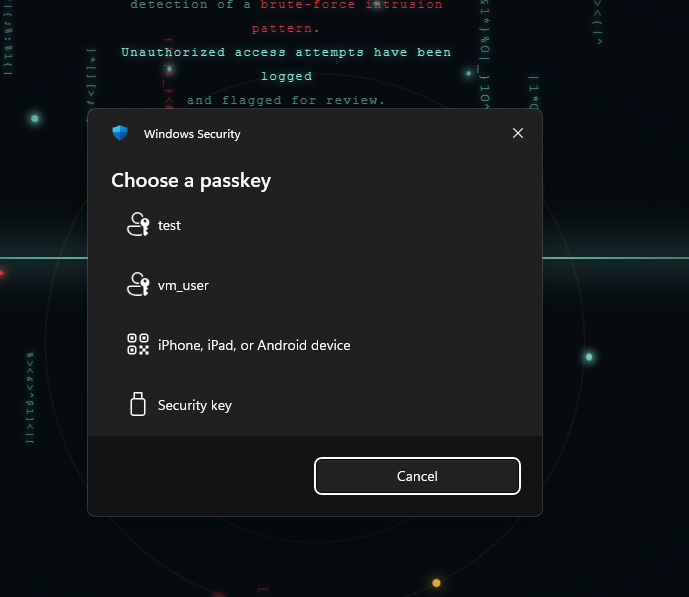}}
\caption{Passkey choice}
\label{FIG:passkey}
\end{figure}

A web-based portal was created for the password reset flow that used the proposed standard to rotate passwords of blocked users. It offered the user to enter the new password, while the username was chosen from the resident credentials of a passkey. Figures \ref{FIG:homepage} shows the homepage while \ref{FIG:passprompt} and \ref{FIG:passkey} show the prompts for password and passkey, respectively.

\section{Results}
\begin{figure}[!htbp]
\centerline{\includegraphics[width=\linewidth]{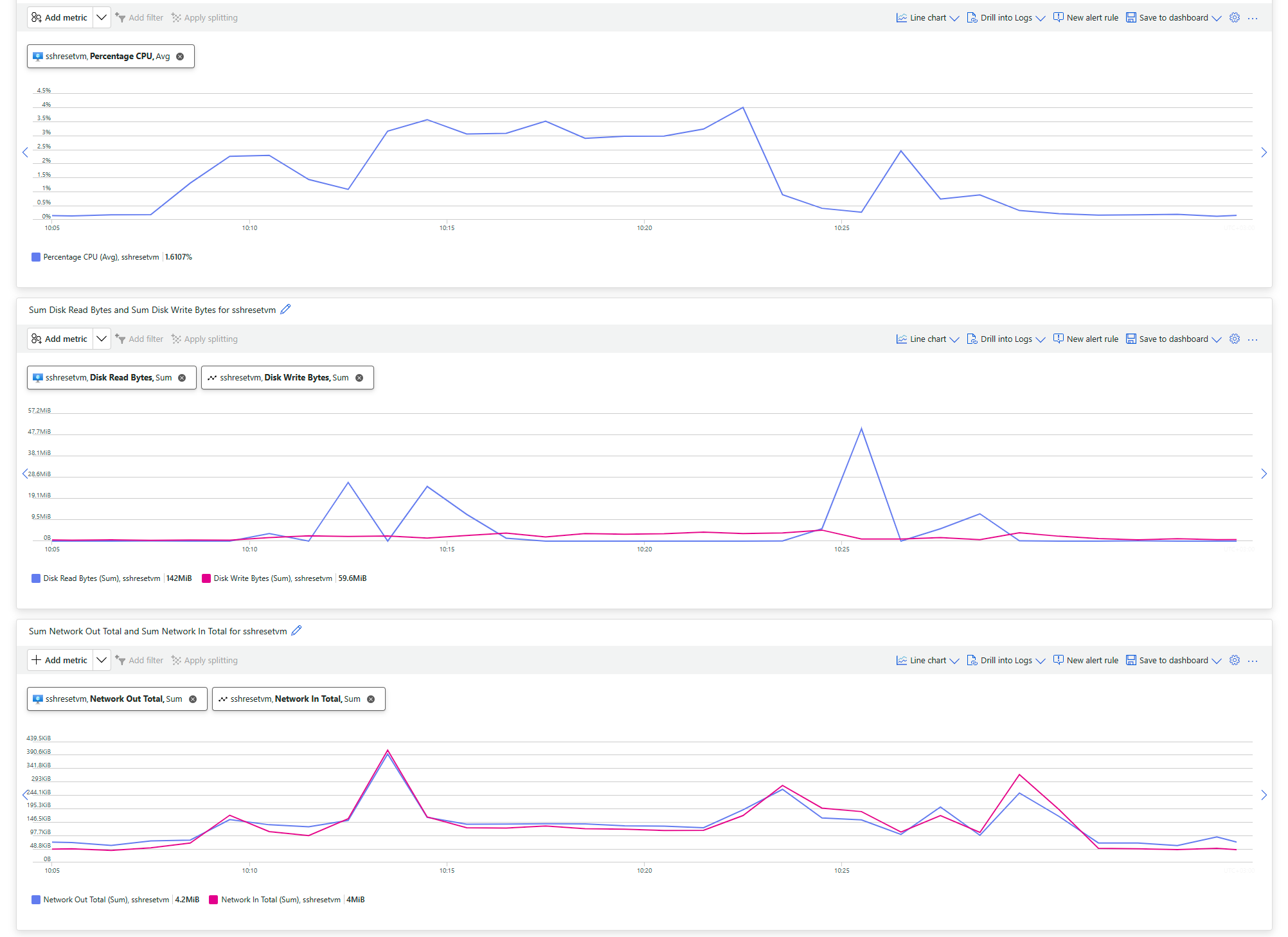}}
\caption{Stats under attack}
\label{FIG:attack}
\end{figure}

The system was able to detect and mitigate brute force attacks on SSH. A control test was done without the proposed system enabled on the same Virtual machine with Hydra. The attack persisted and showed all characteristics of a brute force attack, like high network usage, high disk write for logs, and high CPU usage. Figure \ref{FIG:attack} shows the resource consumption of the VM under the brute force attack.

\begin{figure}[!htbp]
\centerline{\includegraphics[width=\linewidth]{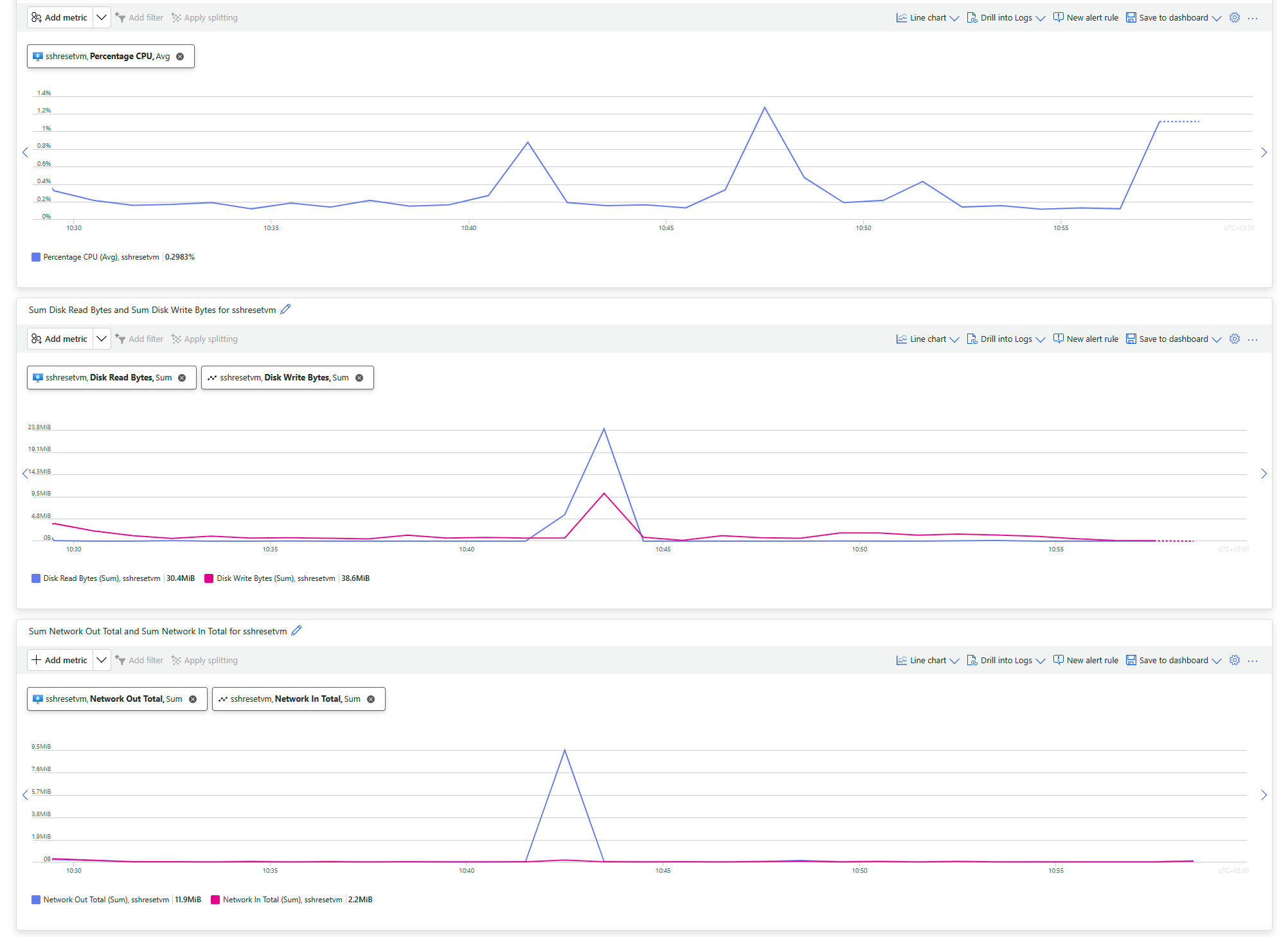}}
\caption{Stats after mitigation}
\label{FIG:mitigated}
\end{figure}

Further, an identical test was performed with the same VM but with the proposed system running. It showed a slight spike in inbound network usage, followed by high disk write, showing the existence of a brute force attack. But it was short-lived, as the system detected it and blocked the user account, which later forced Hydra to exit because password authentication was no longer available. The system had a higher outbound network usage than inbound, unlike the control test, and this can be attributed to the system using a large banner for the blocked accounts, which was sent for all blocked brute-force attempts. The CPU did not spike, unlike the control test, and it mitigated the incoming attack. Figure \ref{FIG:mitigated} shows the resource usage of the VM on mitigation with the proposed system.

\section{Threat model}
The proposed system aims to be more effective than traditional SSH attack detection systems like Fail2Ban and provide a more robust system for account recovery. The threat model of the system includes:
\begin{itemize}
    \item Brute force attacks: The system is aimed to detect and mitigate brute force attacks on the SSH system.
    \item Dictionary attacks: Dictionary attacks are when an attacker attempts to build a personalized or non-personalized wordlist per user and uses multiple passwords from there. The system is able to mitigate such attacks.
    \item Password spray: Password spray attacks involve using a number of known passwords till one succeeds. The system is able to mitigate against such attacks.
    \item Phishing on account recovery flow: Account recovery flow uses a phishing-resistant authentication system and mitigates phishing.
    \item Unexpired sessions: The account recovery flow does not use long-lived sessions, which can be used by attackers.
    \item Cookie theft and session hijack: Cookies and Session cookies used in the account recovery flow cannot be used to authenticate the user or the attacker. Hence, any attack involving them would not result in an account takeover.
    \item CSRF: The account recovery page does not take inputs regularly and instead derives the username from resident credentials. Hence, it cannot be triggered by CSRF attacks.
\end{itemize}

\section{Conclusion}
SSHafe demonstrates to be a solution for fast detection and mitigation of SSH attacks. However, the system may, in future, be combined with other signals from other intrusion detection systems and network and security appliances for better attack detection, not just for SSH, but also for various other types of attacks.

The novel password rotation flow over passkeys may be used for other secret sharing flows as well, not just for account recovery. The password rotation flow aims to be instrumental for sessionless and stateless authenticated secret sharing.

SSHafe has been deployed on real-world cloud environments where it was discovered and attacked by malicious actors and the system has proven to be able to mitigate them, even against real-world adversarial traffic of unknown origin and tooling. The standard proposed in SSHafe would be instrumental in detecting and mitigating various types of attacks in the future, not just SSH.

\section*{Acknowledgment}

This research was conducted within the framework of the Erasmus Mundus Joint Master’s Programme in Applied Cybersecurity (CyberMACS) — Project No. 101082683, co-funded by the European Union under the Erasmus+ MUNDUS Programme.


\bibliography{sn-bibliography}

\end{document}